\documentclass[aps,prl,showpacs,amsmath,twocolumn,amssymb,superscriptaddress,letterpaper]{revtex4}
\usepackage{graphicx,color}
\usepackage{amssymb}   % for math
\usepackage{amsmath}
\usepackage{epstopdf}
\usepackage{natbib}
\usepackage{hyperref}
\usepackage{bm}
\usepackage{xcolor}

\begin{document}
\title{Non-Hermiticity stabilized Majorana zero modes in semiconductor-superconductor nanowires}

\author{Hongchao Liu}
\affiliation{Beijing Academy of Quantum Information Sciences, Beijing 100193, China}
\affiliation{International Center for Quantum Materials, School of Physics, Peking University, Beijing 100871, China}
\author{Ming Lu}
\affiliation{Beijing Academy of Quantum Information Sciences, Beijing 100193, China}

\author{Yijia Wu}
\affiliation{International Center for Quantum Materials, School of Physics, Peking University, Beijing 100871, China}
\author{Jie Liu}
\email{jieliuphy@xjtu.edu.cn}
\affiliation{Department of Applied Physics, School of Science, Xian Jiaotong University, Xian 710049, China}
\author{X. C. Xie}
\affiliation{International Center for Quantum Materials, School of Physics, Peking University, Beijing 100871, China}
\affiliation{CAS Center for Excellence in Topological Quantum Computation,
University of Chinese Academy of Sciences, Beijing 100190, China}

\begin{abstract}
Coupled Majorana zero modes with nonzero energies are generally detrimental to the non-Abelian statistics due to the additional dynamic phase.
Nevertheless, we show that a well-connected lead can introduce a local non-Hermitian dissipation term to shift the energies of the both coupled Majorana modes to zero, and surprisingly turn the coupled Majorana mode far from the lead into a dark Majorana mode with exponentially small dissipation. 
This dark Majorana mode can conquer the drawback of the partially overlapped Majorana zero modes and possess the properties of true Majorana zero mode such as the perfect fractional Josephson effect and the non-Abelian statistics. 
\end{abstract}

\maketitle

{\emph {Introduction}} --- Exotic Majorana zero modes (MZMs) in topological superconducting system have been drawing extensive attention
 during the last decade \cite{kitaev, nayak, Fu, sau, fujimoto, sato, alicea2, lut, oreg, potter, 2DEG1, 2DEG2}.
Since the first signal of MZMs was observed, remarkable experimental progress has been made in various platforms \cite{kou, deng, das1, hao1, hao2, Marcus, perge, Yaz2, Jia, Fes1, Fes2, Fes3, PJJ1, PJJ2}. 
Meanwhile, there're also other possibilities like quasi-MZMs \cite{Jie1, brouwer, Aguado, ChunXiao1, Moore, Wimmer, Aguado2, Klinovaja, Tewari, ABS, ABSM}.
These quasi-MZMs are actually a pair of coupled MZMs separated by a finite distance, and their nonzero energies can lead to undesirable dynamic phase in their time-evolution, making them inappropriate for topological quantum computation (TQC).
Eliminating the influence of quasi-MZMs and distinguishing them from real ones thus become important tasks in Majorana physics \cite{ChunXiao1, Moore, Wimmer, Aguado2, ABS1, ABS2, ABS3, ABSM1, ABS4, ABS5, ABS6, ABS7, ABS8, ABS9, ABS10, ABS12, Zhang1,Zhang2, DongELiu1,DongELiu2}.

Apart from the coupled MZMs, the nanowire has to be connected to leads for transport studies, which introduce non-Hermitian self-energies.
The interplay of non-Hermiticity and topology is predicted to induce many amazing phenomena like the non-Hermitian skin effect and the non-Bloch bulk-boundary correspondence \cite{NH1,NH2,NH3,NH4,NH5,NH6}.
Recent theories suggest that the coupled MZMs could be brought back to exact zero energy with the assistance of the non-Hermitian dissipation from the leads, and the lifetime is also bifurcated \cite{NHM1,NHM2}.
The dissipation is fatal for TQC because it may reduce the lifetime $\tau$ of the coupled MZMs and squeeze up the adiabatic time window $\hbar/\Delta \ll T\ll \tau$ of the braiding process \cite{huang1,Gefen1,adia1}, where $\Delta, T$ are the superconducting gap and the time scale of the braiding operation respectively. 
However, here we show that the local dissipation at one end of the nanowire can counterintuitively prolong the lifetime of the coupled MZM at the other end, thus making the latter more favorable for braiding and TQC. In consideration of the ``dark states'' where dissipation could facilitate decoherence-free states in an unusual manner \cite{dark_np,dark_prl}, we call the coupled MZM stabilized by dissipation a dark Majorana mode (DMM). 

\begin{figure}
\centering
  \includegraphics[width=0.48\textwidth]{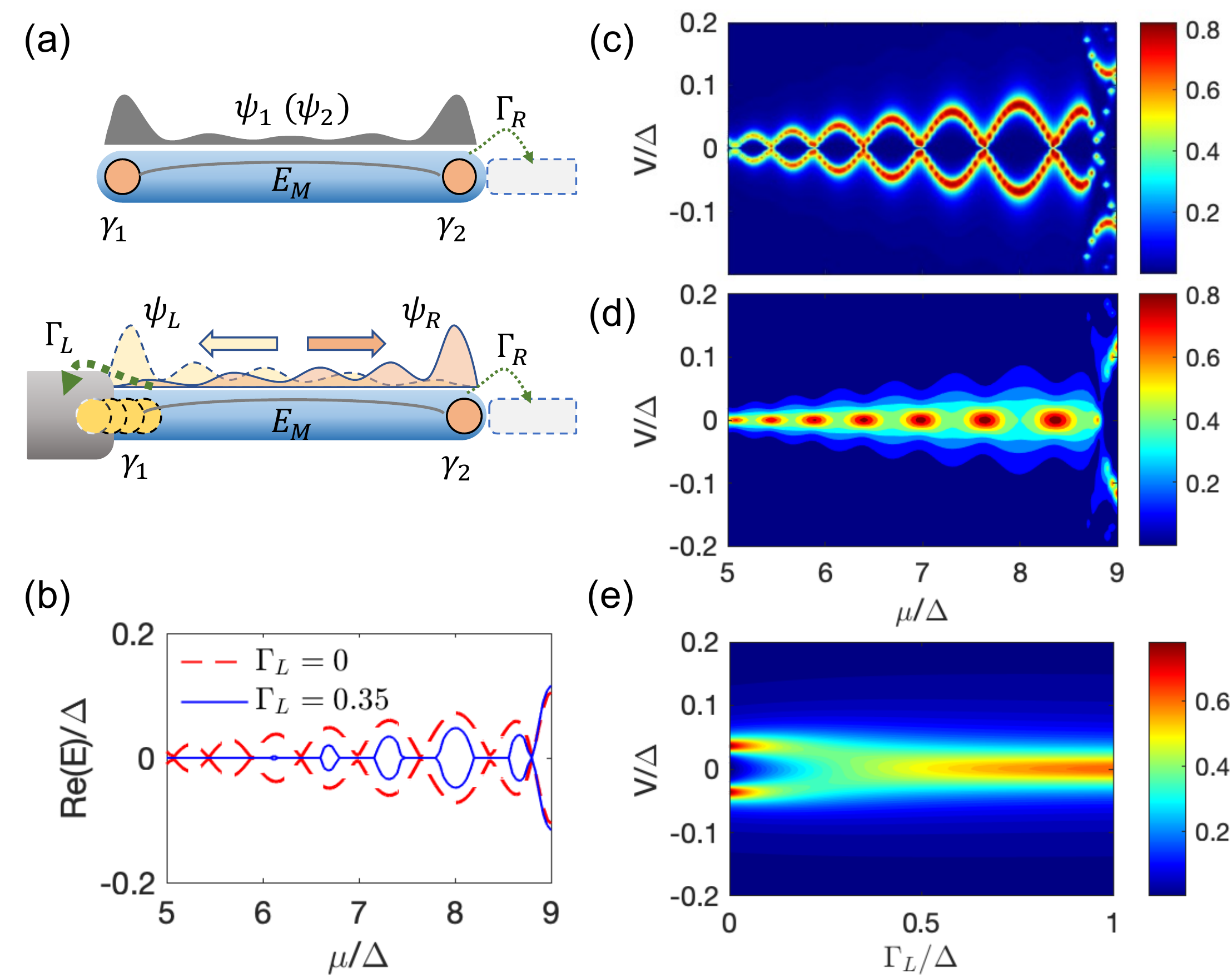}
\caption{
  (a) Schematic diagrams for an isolated nanowire (upper panel) and one strongly coupled with a left lead (lower panel). A right lead is present only when we calculate the differential conductance $dI_R/dV$.
  The coupling between two MZMs is $E_M$ and the i-th lead introduces  dissipation $\Gamma_i$. The wavy curves denote the probability density of the lowest two eigenstates $|\psi_1(x)|, |\psi_2(x)|$. 
  In the upper panel, the two states $|\psi_1(x)|= |\psi_2(x)|$ have a common non-local distribution (dark grey). 
  In the lower panel, the two states are split into spatially separated $|\psi_L(x)|, |\psi_R(x)|$ by the self energy $\Gamma_1$ of the left lead. 
  (b) The real parts of the lowest two eigenenergies for an isolated nanowire (red) and those for one strongly coupled with left lead (blue). Here $\Gamma_R = 0$.  
  (c) $dI_R/dV(2e^2/h)$ versus $V$ and $\mu$, for $\Gamma_L=0$, $\Gamma_R = 0.1\Delta$. 
  (d) $dI_R/dV(2e^2/h)$ versus $V$ and $\mu$, for $\Gamma_L=1.0\Delta$, $\Gamma_R = 0.1\Delta$. 
  (e) $dI_R/dV(2e^2/h)$ versus $V$ and $\Gamma_L$, for $\Gamma_R = 0.1\Delta$, $\mu= 7.5\Delta$.
}
\label{f1}
\end{figure}

The basic structure of our device is shown in the lower panel of Fig. \ref{f1}(a) without the right lead, and the left lead introduces a local non-Hermitian dissipation with two merits. 
Firstly, the coupling between the MZMs is suppressed, so the coupled MZMs are pinned to zero energy and spatially ``polarized'' towards different ends rather than equally-weighted on both ends. 
The energy shift is a non-local effect and could be revealed in transport studies.
Secondly and more importantly, as the  coupled MZM mainly localized on the right end of the nanowire still have a small weight on the left end, the local non-Hermiticity from the left lead can strongly reduce the effective dissipation and the left-end weight of the right-end coupled MZM, making it a DMM.  
Under this non-local effect, the right coupled MZM has exact zero energy and exponentially small dissipation (in the order $10^{-4}\Delta$ or smaller)[Fig. \ref{f2}], which is quite favorable for braiding. 
An additional advantage of the non-Hermiticity is that, since the dissipation of the DMM is also much smaller than the original energy splitting $E_M$ of the isolated nanowire, the condition for adiabatic braiding is much more relaxed. 
We demonstrate that the DMM can preserve all the properties of a true MZM, such as the fractional Josephson effect and the non-Abelian statistics, through both theoretic analysis and numerical simulation.

{\emph{Model}\label{sec:models}---
We use the tight-binding model \cite{Jie1} to describe the quasi-one-dimensional s-wave superconducting nanowire with the Rashba spin-orbit coupling shown in Fig. \ref{f1}(a):
\begin{eqnarray}\label{model}
 H_{q1D}& =& \sum\nolimits_{\mathbf{R},\mathbf{d},\alpha } { -t(\psi _{\mathbf{R} + \mathbf{d},\alpha }^\dagger  \psi _{\mathbf{R},\alpha }  + h.c.) -  \mu \psi _{\mathbf{R},\alpha }^\dagger  \psi _{\mathbf{R},\alpha } } \nonumber \\
&+& \sum\nolimits_{\mathbf{R},\mathbf{d},\alpha ,\beta } { -i{U _R} \psi _{\mathbf{R} + \mathbf{d},\alpha }^\dagger  \hat z \cdot (\vec{\sigma}  \times \mathbf{d})_{\alpha \beta }   \psi _{\mathbf{R},\beta } } \nonumber  \\
 &+& \sum\nolimits_{\mathbf{R},\alpha ,\beta } { \psi _{\mathbf{R}, \alpha }^\dagger V_x (\sigma_x)_{\alpha \beta} \psi _{\mathbf{R},\beta } } \nonumber \\
& + &  \sum\nolimits_{\mathbf{R},\alpha} \Delta \psi _{\mathbf{R},\alpha }^{\dagger} \psi _{\mathbf{R},-\alpha }^{\dagger}+h.c. ,
 \end{eqnarray}
Here $\mathbf{R}$ denotes the lattice sites,
$\mathbf{d}=\mathbf{d_{x}}, \mathbf{d_y}$ denotes the two unit vectors in the $x$ and $y$ directions respectively, $\alpha, \beta$ denotes the spin, $t$ is the hopping amplitude, $\mu$ is the chemical potential, $U_{R}$ is the Rashba coupling strength,
$V_{x}$ is the Zeeman energy caused by an axial magnetic field, and $\Delta$ is the superconducting pairing amplitude. The parameters are set to $\Delta= 0.25 \mathrm{ meV}$, $t=25\Delta$, $V_x=2.5\Delta$ and $U_{R}=2.5\Delta$.
The dimensions of the nanowire are $N_x a \approx 750\mathrm{nm}$, $ N_y a \approx 50 \mathrm{nm}$
($a=10\mathrm{nm}$ is the lattice constant). The superconducting coherence length $\xi_{0} \approx t a/\Delta= 250 \mathrm{nm}$, and thus a pair of coupled MZMs is formed at the ends.

In the presence of the leads, the scattering matrix can be derived using the recursive Green's function method \cite{Jiec}. 
\begin{equation}
  S_{ij}^{\alpha\beta}(E) = -\delta_{i,j} \delta_{\alpha,\beta} + i [\Gamma_i^\alpha]^{1/2} * G^r * [\Gamma_j^\beta]^{1/2}
\end{equation}
where $i,j \in \{L,R\}$ denote two leads, $\alpha,\beta \in \{e,h\}$ denote an electron or a hole, $S_{ij}^{\alpha\beta}$ denotes the scattering amplitude of a $\beta$ particle from lead $j$ to an $\alpha$ particle in lead $i$, $G^r = [E- H_{q1D} - \sum_{i,\alpha} (\Sigma_i^\alpha)^r] $ is the retarded Green's function of the nanowire, $\Gamma_i^\alpha = i[(\Sigma_i^\alpha)^r - (\Sigma_i^\alpha)^a]$ is the linewidth function of $\alpha$ particle in lead $i$, and $(\Sigma_i^\alpha)^{r(a)}$ is the retarded (advanced) self-energy of $\alpha$ particle in lead $i$. In the wide-band limit, $\Gamma_i^e = \Gamma_i^h = \Gamma_i$.

Now we can explore the non-local signature of the non-Hermitian self-energy of the left lead by numerically calculating the differential conductance $dI_R/dV$ in the right lead. The left lead can be well connected to the nanowire to introduce big non-Hermitian dissipation $\Gamma_L$, while the right lead is only used as a probe for tunneling measurement and introduces small $\Gamma_R$. The bias $V$ is applied on the right lead for measurement, while the nanowire and the left lead are grounded, and the current measured from the right lead is \cite{ABS2}
\begin{multline}
  I_R = \frac{e}{h} \int dE [T_\mathrm{LAR} (f_{Re} - f_{Rh}) \\ + (T_\mathrm{e} + T_\mathrm{CAR}) (f_{Re} -f_{L})].
\end{multline}
Here $T_\mathrm{LAR} = |S_{RR}^{eh}|^2, T_\mathrm{e} = |S_{RL}^{ee}|^2, T_\mathrm{CAR} = |S_{RL}^{eh}|^2 $ are the coefficient of the local Andreev reflection (LAR) in the right lead, the electron transmission and the crossed Andreev reflection (CAR) from the left lead to the right lead, respectively; $f_{Re} = 1-f_{Rh} = f(E - e V) $ means the Fermi distribution function of electrons/holes in the right lead and $f_{L} = f(E) = [1+ \exp(E/k_B T)]^{-1}$ means the Fermi distribution function of elections/holes in the left lead, which is at zero bias. If we consider zero temperature Fermi distribution function $f(E) = \theta(-E)$ for simplicity, then the differential conductance is given by
\begin{equation}\label{eq_zbp1}
  \frac{dI_R}{dV} = \left( |S_{RR}^{eh}|^2 + \frac{ |S_{RL}^{ee}|^2 + |S_{RL}^{eh}|^2 }{2} \right) \frac{2e^2}{h} 
\end{equation}
where all the scattering matrix elements are evaluated at $E=eV$.

{\emph {Non-Hermiticity stabilized zero bias peak}} ---
First we only weakly connect the nanowire to the right lead for conductance measurement, and don't connect the left lead [Fig. \ref{f1}(a), upper panel]. 
Due to the finite size effect, the eigenstates are the coupled MZMs which are superpositions of the true MZMs $\gamma_1, \gamma_2$, and the eigenenergies are generally non-zero. Therefore, the conductance peak will split away from zero energy [Fig. \ref{f1}(c)]. 

Then we connect the nanowire to a left lead, which introduces a strong dissipation $\Gamma_L$ [Fig. \ref{f1}(a), lower panel]. 
Now the position of the peak of $dI_R/dV$ will be significantly suppressed towards zero energy [Fig. \ref{f1}(d)], revealing a non-local pinning effect by $\Gamma_L$ of the left lead. 
To understand this energy shift, consider a minimal non-Hermitian Hamiltonian in the Majorana basis \cite{NHM1}, 
  \begin{eqnarray}\label{model2}
  H_{\mathrm{NH}}=\begin{pmatrix}
  -i\Gamma_L &-i E_M \\iE_M &-i\Gamma_R
  \end{pmatrix},
  \end{eqnarray}
where $E_M$ is the coupling of $\gamma_1,\gamma_2$, and $\Gamma_i$ is the dissipation on $\gamma_i$. In our model the MZMs $\gamma_1,\gamma_2$ are at the ends of the nanowire, while $\Gamma_L$ and $\Gamma_R$ are from left and right leads, respectively \cite{supple}. 
The eigenvalues are $E_{\pm}=-i(\Gamma_L+\Gamma_R)/2\pm [E_M^2- (\Gamma_L-\Gamma_R)^2/4]^{1/2} $.
If both leads are weakly connected, $\Gamma_L-\Gamma_R \ll E_M$ and the two levels only slightly deviate from $ \pm E_M$.
On the contrary, when only the left lead is well connected, the asymmetric dissipation is strong $\Gamma_L-\Gamma_R \gg E_M$ and both levels will be pinned to zero. This heuristic picture is qualitatively consistent with the numerical results shown in Fig. \ref{f1}(b).

To further reveal this energy shift, we can keep the right lead weakly connected, while modulate the coupling strength between the left lead and the nanowire. 
Because the right lead is not changed, the energy shift of the peak of $dI_R/dV$ due to the change of $\Gamma_L$ is a non-local behavior and can be used for distinguishing coupled MZMs from local fermionic states. 
The energy shift, shown in Fig. \ref{f1}(e) numerically, is also consistent with the analytical result
$\frac{dI_R}{dV} =\frac{1} {|Z|^2} \left( \Gamma_R^2(E^2+\Gamma_L^2)+E_M^2\Gamma_L\Gamma_R \right)\frac{2e^2}{h} $
in Refs. [\onlinecite{Jiec,nilsson}], with $Z = E_M^2 - (E+i\Gamma_L) (E+ i\Gamma_R) $, $E=eV$ is the energy of the particle. The eigenenergies $E_\pm$ of the minimal model Eq. (\ref{model2}) are just the poles of $\frac{dI_R}{dV}$. A straightforward derivation \cite{supple} shows that the first term of $\frac{dI_R}{dV}$ is the contribution of LAR, and the second term is the contribution of non-local electron transmission and the CAR, corresponding to Eq. (\ref{eq_zbp1}). For the ideal long nanowire case, $E_M = 0$ and the zero bias peak (ZBP) only contains the perfectly quantized LAR term $G^\mathrm{ideal} = G_\mathrm{LAR}^\mathrm{ideal} = \frac{2e^2}{h}$. In realistic finite size nanowires, Ref. [\onlinecite{Jiec,nilsson}] have focused on weak connection $\Gamma_L, \Gamma_R\ll E_M$ and found that the peaks lie at $\mathrm{Re} E_\pm \approx \pm E_M$. In contrast with this splitting behavior, here we find that if the nanowire is in well connection with one lead but in weak connection with the other lead, $\Gamma_L-\Gamma_R \gg E_M$, $E_\pm$ become purely imaginary, and the peaks will shift to zero and merge. After being pinned to zero energy, the ZBP is different from quantized $\frac{2e^2}{h}$ in the ideal long nanowire case due to non-zero $E_M$,
$G^\mathrm{NH} = \frac{\Gamma_L \Gamma_R }{E_M^2+\Gamma_L\Gamma_R} \frac{2e^2}{h}.$
This ZBP is the sum of a non-quantized LAR term $ G_\mathrm{LAR}^\mathrm{NH} = \frac{(\Gamma_L \Gamma_R)^2 }{(E_M^2+\Gamma_L\Gamma_R)^2} \frac{2e^2}{h}$ and two non-zero non-local term $G_\mathrm{e}^\mathrm{NH} = G_\mathrm{CAR}^\mathrm{NH} = \frac{E_M^2 \Gamma_L \Gamma_R }{2(E_M^2+\Gamma_L\Gamma_R)^2} \frac{2e^2}{h}$.  Although a quantized ZBP is present for $\Gamma_L\Gamma_R\gg E_M^2$, the ZBP in experiment may oscillate with $\mu$, which changes $E_M$. 
Such fluctuation may explain the instability of quantized ZBP observed in recent experiments\cite{Zhang1,Zhang2,IronBase1,IronBase2}.

\begin{figure}
\centering
  \includegraphics[width=0.48\textwidth]{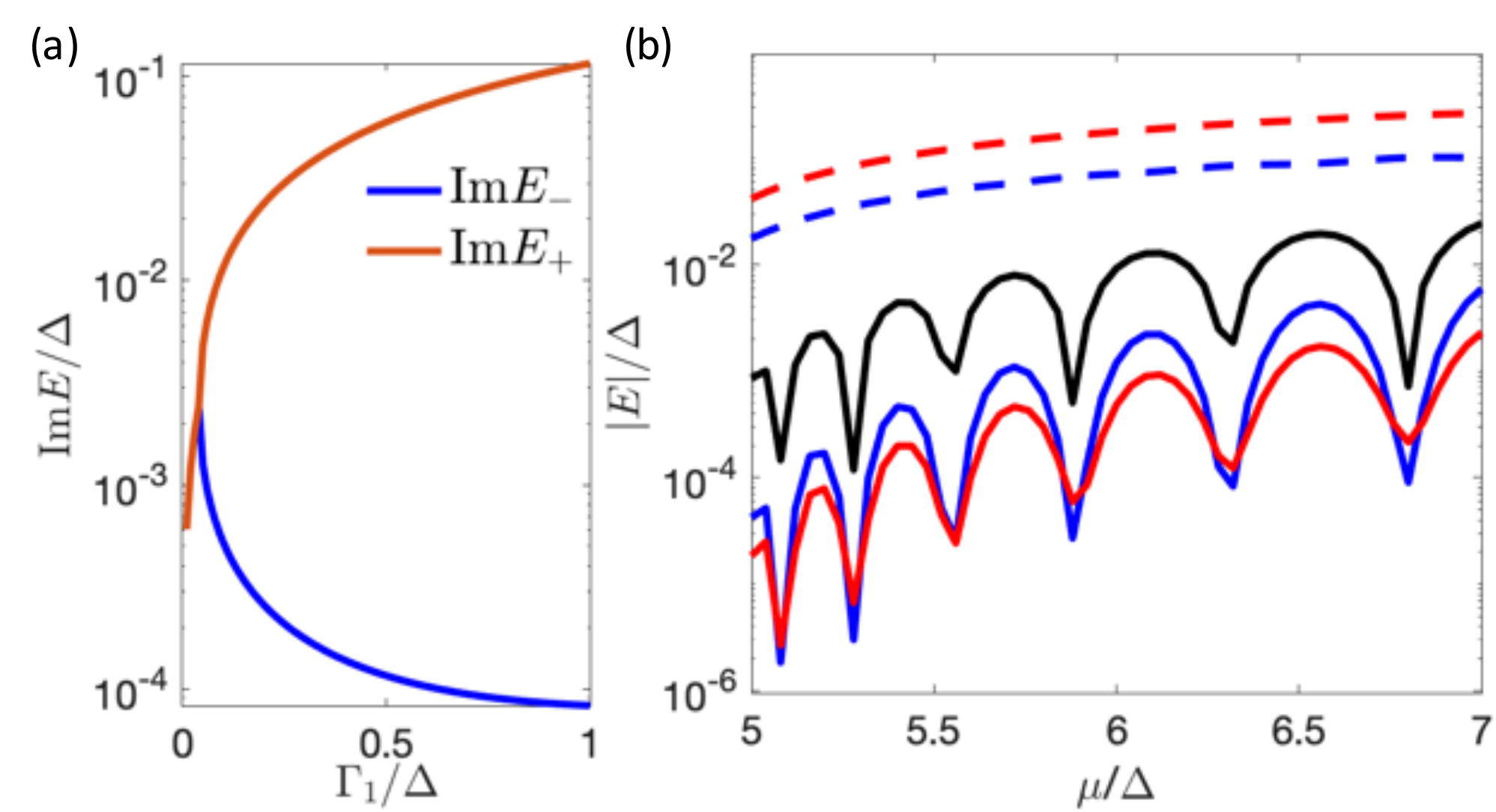}
\caption{
The stability of the DMM. (a) The decay rate of lowest two levels versus the self-energy $\Gamma_L$, with $\Gamma_R=0$, $\mu=6\Delta$. % We estimate $E_M\sim 10^{-3}\Delta$ from our calculation of the isolated nanowire [black line in (b)].
(b) The moduli of the two levels with $\Gamma_L=\Gamma_R=0$ (black, $|E_+|=|E_-|=E_M $), $\Gamma_L=0.5\Delta, \Gamma_R=0$ (blue, dashed for $|E_+|$ and solid for $|E_-|$), and $\Gamma_L=\Delta, \Gamma_R=0$ (red, dashed for $|E_+|$ and solid for $|E_-|$).
}
\label{f2}
\end{figure}

{\emph {Non-Hermiticity tuned perfect DMM}} ---We have shown that the finite hybridization strength can be suppressed to zero through the asymmetric dissipation.
A natural question is whether these dissipation-induced MZMs are stable enough.
Interestingly, Eq. (\ref{model2}) gives the relation
$\operatorname{Im} E_+ \cdot \operatorname{Im} E_-=E_M^2$ for $\Gamma_L>2E_M$ and $\Gamma_R=0$. 
Therefore, by simply keeping the left lead in well connection but detaching the right lead, $\Gamma_L$ increases $\operatorname{Im} E_+ = \Gamma_L/2 + [\Gamma_L^2/4 - E_M^2]^{1/2}$ and decreases $\operatorname{Im} E_-$ monotonously, making the latter a stable mode. The eigenstates corresponding to $E_\pm$ are
are $\psi_{L,R} = \frac{1}{\sqrt{2}}(\sqrt{1\pm \sqrt{ 1- E_M^2 / \Gamma_0^2 } } ,\sqrt{1\mp \sqrt{ 1- E_M^2 / \Gamma_0^2 } })^T$ in the Majorana basis, where the asymmetric dissipation term $\Gamma_0\equiv (\Gamma_L-\Gamma_R)/2$ suppresses the overlap between $\psi_{L,R}$ and pushes them to the left (right) end. 
In the case $\Gamma_L \gg E_M, \Gamma_R=0$, we have $\psi_R \approx (0,1)^T = \gamma_2$, polarized to the right end and converges to $\gamma_2$ by $\Gamma_L$ in this limit, as shown in the lower panel of Fig. \ref{f1}(a). 
This also explains why the stronger left dissipation lessens the dissipation on the $\psi_R$: it suppresses (enhances) the weight of $\gamma_1 (\gamma_2)$ in $\psi_R$, and because the true MZM $\gamma_2$ is locally immune from the left lead, $\psi_R$ feels less dissipation overall.

\begin{figure}
  \centering
    \includegraphics[width=0.48\textwidth]{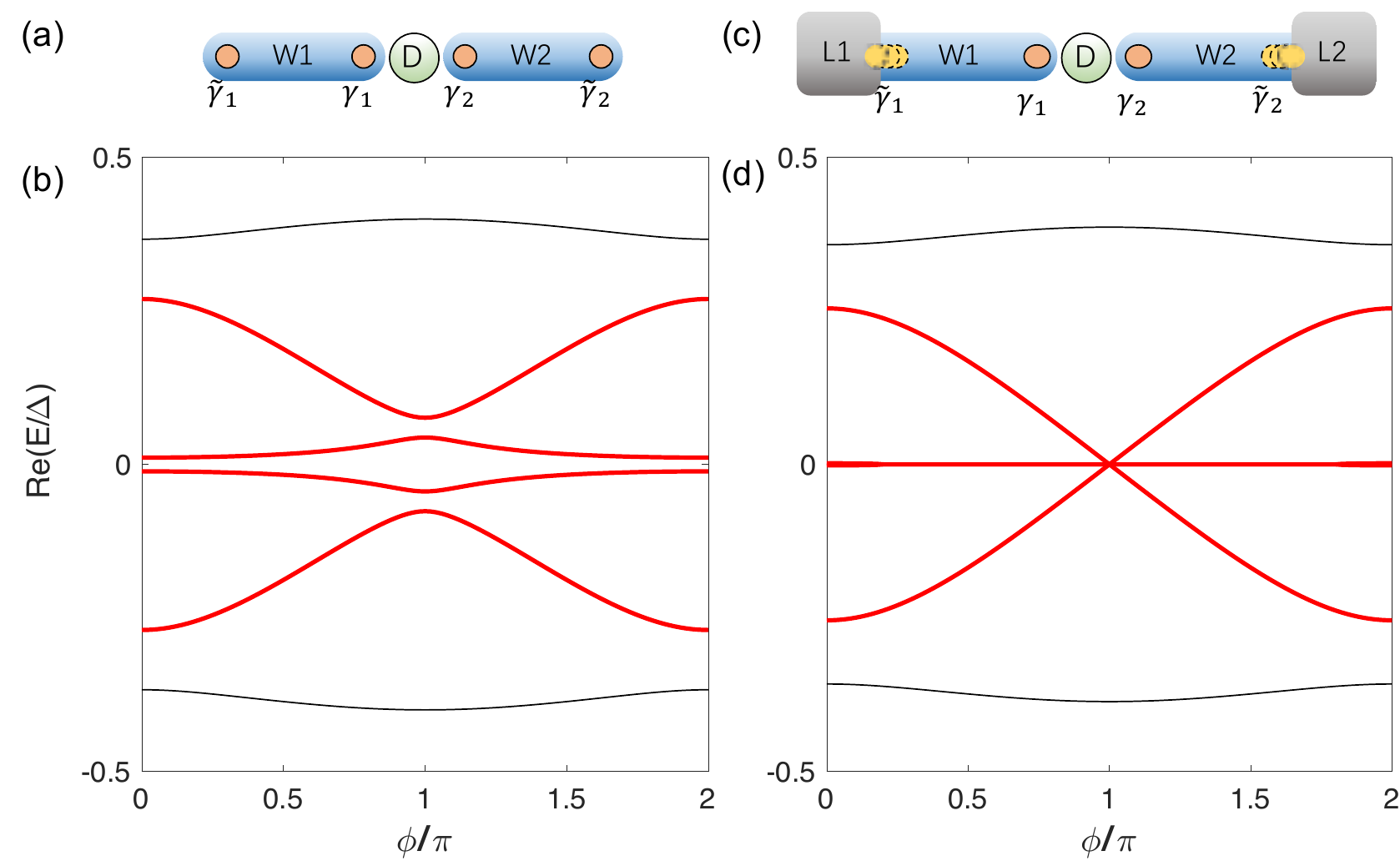}
  \caption{
    Topological Josephson junction.
    (a) A junction made by two nanowires (blue, with $N_x a=750\mathrm{nm}$) and a quantum dot (green) in the center. The finite size effect causes the coupling of four MZMs $ \gamma_1, \gamma_2,\tilde{\gamma_1},\tilde{\gamma_2}$.
    (b) The Andreev bound states versus the phase difference for (a), showing anticrossing and $2\pi$ period.
    (c) Same as (a) except the two outer ends of the system are connected to leads. The leads introduce $\Gamma_L=\Delta$ for the MZMs $\tilde{\gamma}_1, \tilde{\gamma}_2$ far from the quantum dot and $\Gamma_R=0$ for the MZMs $\gamma_1,\gamma_2$ near the quantum dot.
    (d) The Andreev bound states versus the phase difference for (c), showing crossing and $4\pi$ period.
  }
  \label{f3}
  \end{figure}

This heuristic result is consistent with our numerical one shown in Fig. \ref{f2}. As the coupling between the lead and the nanowire increases to $\Gamma_L\gg E_M$, the dissipation for  $\psi_L$ also increases to $\operatorname{Im} E_+\gg E_M$, but that for $\psi_R$ decreases to $\operatorname{Im} E_- \ll E_M$. Fig. \ref{f2} (b) shows the moduli of the eigenvalues versus the chemical potential $\mu$ with $N_x a =750\mathrm{nm}$. Without the dissipation, the levels are real values and $E_M \sim 10^{-3}\Delta$ (black line). While if we increase $\Gamma_L$ to $0.5\Delta$, the energies become purely imaginary with $\operatorname{Im} E_-\sim 10^{-4} \Delta$ (blue line).
If we further increase $\Gamma_L$ to $1.0 \Delta$, which is the same as that in Fig. \ref{f1}(d) and easily accessible in experiments, then $\operatorname{Im} E_-$ further decreases, corresponding to a long lifetime of the coupled MZM $\psi_R$. 
Therefore the coupled MZM $\psi_R$ is a perfect DMM, indicating that it has zero energy and exponentially small dissipation, and approximates to the true MZM $\gamma_2$. Armed with these properties, the DMM is expected to show most of the properties of a true MZM.

Indeed, we found this idea can be employed in a topological Josephson junction as shown in Fig. \ref{f3}(a).
It is well known that the MZMs may induce fractional Josephson effect, but the effect is vulnerable to the finite size effect of the nanowire \cite{Jie2}. In Fig. \ref{f3}(b) we show the energy level of Andreev bound states versus the phase difference of two nanowires. The energy levels anticross each other due to the finite size effect, restoring the system to be $2\pi$ periodic.
On the contrary, if we add two normal leads at the outer ends as shown in Fig. \ref{f3}(c), then two DMMs can be created near the quantum dot.
These two DMMs will couple with each other and not be affected by the other two short-lived quasi MZMs at the outer ends, so their energy-phase relation shows a perfect $4\pi$ periodic behavior [Fig. \ref{f3}(d)].

\begin{figure}
  \centering
    \includegraphics[width=0.48\textwidth]{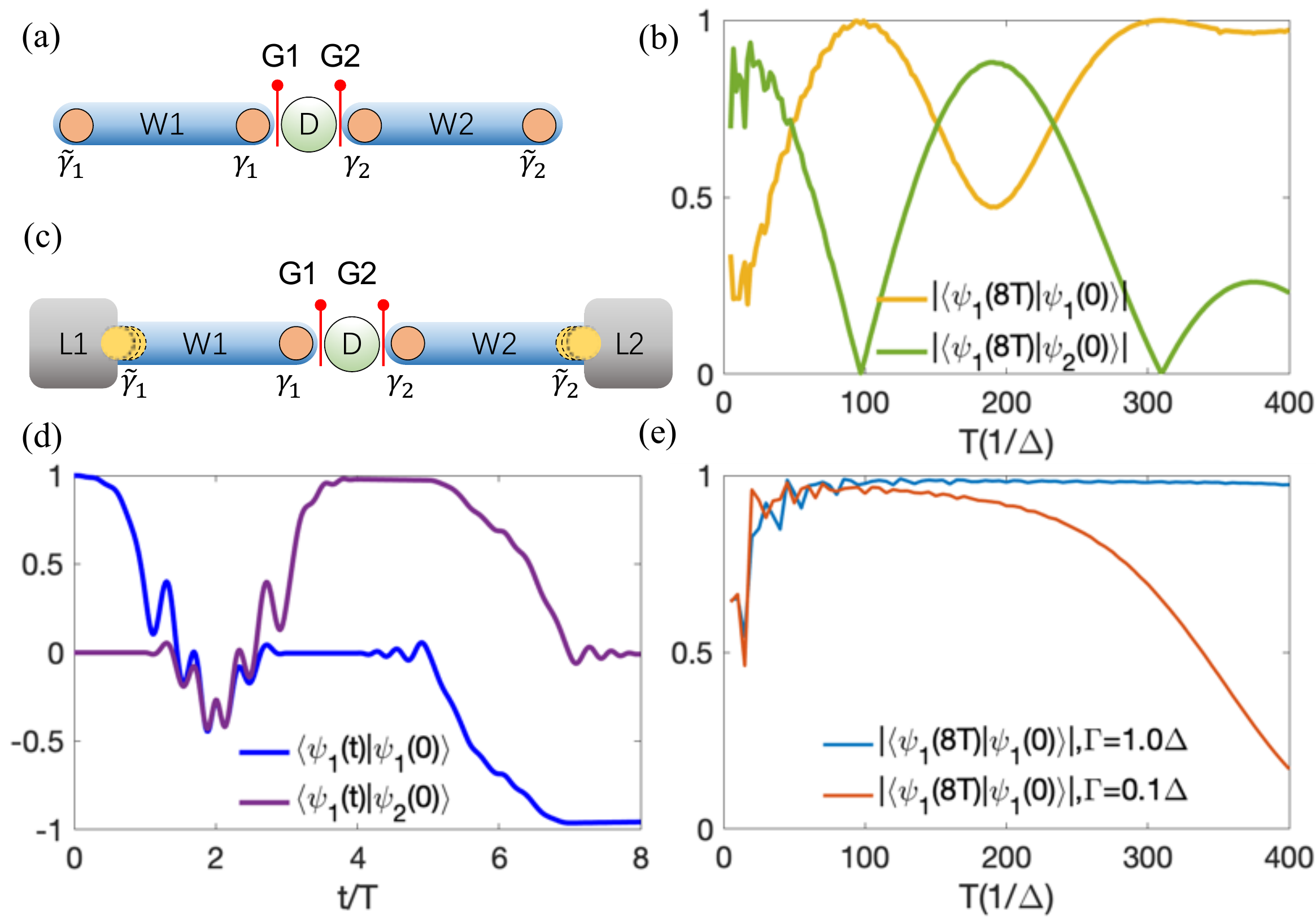}
  \caption{
    Non-Abelian braiding of the DMMs. 
    (a) A device for braiding two MZMs $ \gamma_1,\gamma_2$. W1, W2 are two nanowires, G1, G2 are the electrical gates between the nanowires and the quantum dot D. 
    (b) The braiding fidelity versus $T$ for (a), where $\psi_{1(2)} (0) = (\gamma_1\pm i\tilde \gamma_1)/\sqrt{2} $.
    (c) Same as (a) but W1, W2 are connected with leads L1, L2.
    (d) The time evolution of the wavefunctions $ \psi_{1(2)} $ for (c), where $\psi_{1(2)} (0) = \gamma_{1(2)} $ and time scale $T=100/\Delta$.
    (e) The braiding fidelity versus $T$ for (c), where $\psi_{1(2)} (0) = \gamma_{1(2)} $. $\Gamma$ is the dissipation from the leads L1, L2.
  }
  \label{f4}
  \end{figure}

{\emph {Non-Abelian statistics of DMM}} ---
We have shown that a local dissipation term on one end can produce a perfect DMM on the other end; however, the possible non-Abelian statistics of the DMMs remains to be investigated.
The non-Abelian statistics of MZMs is induced by a non-trivial geometric phase of $\pi$ in the braiding process, where two MZMs are spatially swapped. 
The braiding operator $B(\gamma_i,\gamma_j) = \exp(\frac{\pi}{4}\gamma_i\gamma_j)$
transforms the MZMs as $\gamma_i\rightarrow \gamma_j$ and $\gamma_j\rightarrow -\gamma_i$ \cite{Ivanov}.
If the nanowires W1, W2 in Fig. \ref{f4}(a) are long enough to suppress the coupling energy $E_M$ between the MZM near the dot $\gamma_i$ and that far from the dot $\tilde\gamma_i$, the non-local fermions for the left nanowire W1 will be $\psi_{1(2)}(0) = (\tilde \gamma_1\pm i\gamma_1)/\sqrt{2}$.
If $\gamma_1$ and $\gamma_2$ are braided twice in succession, the wavefunctions for W1 will be
$\psi_{1} (8T) =  (\tilde \gamma_1- i\gamma_1)/\sqrt{2}\equiv \psi_2(0)$. Here the time duration for the braiding operation is $4T$ and the detailed braiding protocol could be found in \cite{Jienew, supple}. 
Then in this case, the fidelity of the braiding can be defined as $|\langle\psi_1 (8T)|\psi_2(0)\rangle| $, which is expected to reach unity for a successful braiding. 
Because of the finite size of W1, W2, $E_M$ will generate a dynamic phase for large $T$.
To see this, the wavefunction evolution $|\psi_{1(2)} (t)\rangle=U(t)|\psi_{1(2)} (0)\rangle$ is calculated, with
$U(t) = \hat{T} \exp[i\int_0^{t} \mathrm{d}\tau H_s(\tau)]$ the time-evolution operator, $H_s$ the Hamiltonian of the system and $\hat{T}$ the time-ordering operator \cite{Jie3, Dirac}.
For $E_M \approx 10^{-3}\Delta$, due to the dynamic phase, $\psi_1 (8T)$ tends to fall back to $\psi_1 (0)$ and the fidelity is obviously lower than unity at large $T$, as shown in Fig. \ref{f4}(b). 
For small $T$, the fidelity also fails to reach our expectations, probably because of the involvement of supragap state. 
Therefore $E_M \approx 10^{-3}\Delta$ is already too large to implement the adiabatic braiding condition $\hbar/\Delta \ll T \ll \hbar/E_M $ \cite{adia1}.

In contrast, if we connect W1, W2 to leads L1, L2 [Fig. \ref{f4}(c)], the dynamic phase vanishes 
because under the large non-Hermitian self-energy of the leads, the DMMs $\psi_{W1,R} = \gamma_1, \psi_{W2,L} = \gamma_2$ are pinned at exact zero energy.
In addition, the dissipation introduced by L1, L2 only acts as an identity background $\Gamma_{s}$ during the braiding, thus the braiding operator would be $\tilde{B}(\gamma_i,\gamma_j) = \exp(\frac{\pi}{4}\gamma_i\gamma_j -\int \Gamma_{s} dt)$ and the non-Abelian geometric phase of $\pi$ remains unchanged \cite{supple}.
This is confirmed by the numerically calculated evolution shown in Fig. \ref{f4}(d), with the initial states $\psi_{1(2)} (0) = \gamma_{1(2)} $. 
The wave function $\psi_1(t)$ after braiding once is $ \psi_1 (4T) \propto \psi_2(0)$, and after braiding twice is $ \psi_1 (8T) \propto -\psi_1 (0)$, so the geometric phase is the same as that based on the true MZMs.
Now the fidelity of the braiding should be redefined as $|\langle\psi_1 (8T)|\psi_1(0)\rangle |$. 
As shown in Fig. \ref{f4}(e), larger dissipation $\Gamma$ from the lead can improve the fidelity in a wider region of $T$, because this dissipation suppresses $\operatorname{Im} E_- $ and stabilizes the DMM. 
Compared with the previous case $E_M \approx 10^{-3}\Delta$, here the dissipation on the DMM can be suppressed down to $ \operatorname{Im} E_- \approx 10^{-5}\Delta$, and the energy for DMMs is $E_M =0$, so the condition for adiabatically braiding is significantly relaxed to $\hbar/\Delta \ll T \ll \hbar/\operatorname{Im} E_-$ for non-Abelian braiding based on DMMs.

{\emph {Discussion and Conclusion} ---}
With the assistance of a local dissipation term, we show that a perfect DMM can be prepared in a short
 semiconductor-superconductor nanowire.
These DMMs have their energy pinned to zero with long lifetime and preserve the non-Abelian statistics of MZMs quite well. 
In reality, the quasi-MZMs can emerge due to the inhomogeneous potential at the interface. 
Although they can even persist in a trivial phase, they are still partially separated and can be viewed as a pair of coupled MZMs \cite{Wimmer}. Hence our proposal also applies for the case of quasi-MZMs.
When they masquerade the true MZMs in the transport studies, one of them is probably transformed into a DMM because of the dissipation of the lead. 
Therefore, these seemingly-trivial quasi-MZMs could also become candidates for TQC. 
Surprisingly, though dephasing usually destroys the coherence of quantum qubits, our work suggests that local dephasing may even assist the TQC because it can induce DMM non-locally in the far end of the nanowire with exact zero energy and exponentially small dephasing, greatly relaxing the time scale for adiabatically braiding.
Finally, we'd like to point out that aside from the connection with normal leads, there are many other physical ways to introduce the dissipations,
such as the fluctuation of superconducting phase, external time-dependent driving forces, and the environmental modes.  In principle, these dissipation terms arising from different
sources may work in the same way. Therefore, we expect that these different dissipation terms may provide more experimental platforms supporting DMMs.

{\emph{Acknowledgement}} --- This work is financially supported by NSFC (Grants No. 11974271) and NBRPC (Grants No. 2017YFA0303301, and No. 2019YFA0308403).

\bibliography{Non-hermitian}

\end{document}